\documentclass[vecphys]{svmult}

\usepackage{makeidx}         
\usepackage{graphicx}        
\usepackage{multicol}        
\usepackage[bottom]{footmisc}

\makeindex             

\begin{document}

\title*{Invariant manifolds as building blocks for the formation of
spiral arms and rings in barred galaxies}
\titlerunning{Formation of spiral arms and rings in barred galaxies}
\author{M. Romero-G\'omez\inst{1}\and E. Athanassoula\inst{1}\and J.J. 
Masdemont\inst{2}\and C. Garc\'{i}a-G\'omez\inst{3}}
\institute{L.A.M., Observatoire Astronomique de Marseille Provence,
2 Place Le Verrier, 13248 Marseille (France)
\texttt{merce.romerogomez@oamp.fr // lia@oamp.fr}\and 
I.E.E.C \& Dep. Mat. Aplicada I, Universitat Polit\`ecnica de
Catalunya, Diagonal 647, 08028 Barcelona (Spain)
\texttt{josep@barquins.upc.edu}\and
D.E.I.M., Universitat Rovira i Virgili, Campus Sescelades, 
Avd. dels Pa\"{\i}sos Catalans 26, 43007 Tarragona (Spain)
\texttt{carlos.garciag@urv.cat}
}
\authorrunning{Romero-G\'omez et al.}

\maketitle

We propose a theory to explain the formation of spiral arms and of
all types of outer rings in barred galaxies, extending and applying
the technique used in celestial mechanics to compute transfer orbits. Thus,
our theory is based on the chaotic orbital motion driven by the invariant
manifolds associated to the periodic orbits around the hyperbolic equilibrium
points. In particular, spiral arms and outer rings are related to the
presence of heteroclinic or homoclinic orbits. Thus, $R_1$ rings are associated
to the presence of heteroclinic orbits, while $R_1R_2$ rings are associated 
to the presence of homoclinic orbits. Spiral arms and $R_2$ rings, however, 
appear when there exist neither heteroclinic nor homoclinic orbits. We 
examine the parameter space of three realistic, yet simple, barred galaxy 
models and discuss the formation of the different morphologies according to the
properties of the galaxy model. The different morphologies arise from
differences in the dynamical parameters of the galaxy.

\section{Introduction}
\label{sec:intro}
\index{Introduction}
Bars are a very common feature of disc galaxies. In a sample of $186$
spirals, $56\%$ of the galaxies in the near infrared are strongly barred, 
while an additional $16\%$ are weakly barred \cite{esk00}. A large fraction 
of barred galaxies show two clearly defined spiral arms \cite{elm82}, 
departing from the end of the bar. This is the case for instance in 
NGC~1300, NGC~1365 and NGC~7552. Spiral arms are believed to be density 
waves in a disc galaxy \cite{lind63}. In \cite{too69}, Toomre found that 
the spiral waves propagate towards the principal Lindblad resonances of the 
galaxy, where they damp down, and thus concludes that long-lived spirals need 
some replenishment. Danby argued that orbits in the gravitational 
potential of a bar play an important role in the formation of arms \cite{dan65} and 
Kaufmann \& Contopoulos argue that, in self-consistent models for
three real barred spiral galaxies, spiral arms are supported also by 
chaotic orbits \cite{kau96}. 
 
The origin of rings has been studied by Schwarz who calculated the response 
of a gaseous disc galaxy to a bar perturbation \cite{sch81,sch84,sch85}. 
He proposed that ring-like patterns are associated to the principal orbital 
resonances, namely ILR (Inner Lindblad Resonance), CR (Corotation Resonance), 
and OLR (Outer Lindblad Resonance). There are different types of outer rings. 
Buta classified them according to the relative orientation of 
the ring and bar major axes \cite{but95}. If these two are perpendicular, the outer 
ring is classified as $R_1$. If they are parallel, the outer ring is 
classified as $R_2$. Finally, if both types of rings are present in the 
galaxy, the outer ring is classified as $R_1R_2$.

In Romero-G\'omez {\it et al.} \cite{rom06,rom07}, we note that spiral
arms and rings emanate from the ends of the bar and we propose that rings 
and spiral arms are the result of the orbital motion driven by the invariant 
manifolds associated to the Lyapunov periodic orbits around the unstable 
equilibrium points. In Romero-G\'omez {\it et al.} \cite{rom06}, we fix a 
barred galaxy potential and we study the dynamics around the unstable 
equilibrium points. We give a detailed definition of the invariant manifolds 
associated to a Lyapunov periodic orbit. For the model considered, the 
invariant manifolds delineate well the loci of an $rR_1$ ring structure, i.e. 
a structure with an inner ring (r) and an outer ring of the type $R_1$. 
In Romero-G\'omez {\it et al.} \cite{rom07}, we construct families of 
models based on simple, yet realistic, barred galaxy potentials. In each 
family, we vary one of the free parameters of the potential and keep the 
remaining fixed. For each model, we numerically compute the orbital structure 
associated to the invariant manifolds. In this way, we are able to study the 
influence of each model parameter on the global morphologies delineated by 
the invariant manifolds.

Voglis, Stavropoulos \& Kalapotharakos study the chaotic motion present in 
self-consistent models of both rotating and non-rotating galaxies, concluding 
that rotating models are characterised by larger fractions of mass in chaotic 
motion \cite{vog06a}. Patsis argues that the spiral arms of NGC~4314 are due 
to chaotic orbits and, to show it, he computes families of orbits with initial 
conditions near the unstable equilibrium points \cite{pat06}. Voglis, 
Tsoutsis \& Efthymiopoulos \cite{vog06b} reproduce a spiral pattern found 
in a self-consistent simulation using the apocentric invariant manifolds of 
the short-period family of unstable periodic orbits. They give the angular 
position of the apocentres, which is where they state the stars spend a large 
part of their radial period, as a soliton-type solution of the Sine-Gordon 
equation. 

In Sect.~\ref{sec:mod}, we first present the galactic models used in the 
computations and the equations of motion. In Sect.~\ref{sec:inv}, we give a 
brief description of the dynamics around the unstable equilibrium points and
the role the invariant manifolds play in the transfer of matter around
the galaxy. In Sect.~\ref{sec:res}, we show the different morphologies that 
result from the computations. 

\section{Description of the model and equations of motion}
\label{sec:mod}
\index{Description of the model and equations of motion}
In this section, we first describe the bar models used in the computations
by giving the density distributions, or the potentials used. We then write
the equation of motion and we define the effective potential and Jacobi
constant.

\subsection{Description of the model}
\index{Description of the model}
We use three different models, all three consisting of the superposition of an 
axisymmetric component and another bar-like. Our first model is that of Athanassoula 
\cite{ath92}. The axisymmetric component is composed of a disc, modelled as a 
Kuzmin-Toomre disc \cite{kuz56,too63} of surface density $\sigma(r)$:
\begin{equation}\label{eq:kuz}
\sigma(r) = \frac{V_d^2}{2\pi r_d}\left(1+\frac{r^2}{r_d^2}\right)^{-3/2}, 
\end{equation}
and a spheroid modelled by a density distribution of the form $\rho(r)$:
\begin{equation}\label{eq:sph} 
\qquad \rho(r)=\rho_b\left(1+\frac{r^2}{r_b^2}\right)^{-3/2}.
\end{equation}
The parameters $V_d$ and $r_d$ set the scales of the disc velocities and radii, 
respectively, and  $\rho_b$ and $r_b$ determine the concentration 
and scale-length of the spheroid. 

Our bar potential is described by a Ferrers ellipsoid \cite{fer77} 
whose density distribution is:
\begin{equation}
\left\{\begin{array}{lr}
\rho_0(1-m^2)^n & m\le 1\\
 0 & m\ge 1,
\end{array}\right.
\label{eq:Ferden}
\end{equation}
where $m^2=x^2/a^2+y^2/b^2$. The values of $a$ and $b$ determine the shape of 
the bar, $a$ being the length of the semi-major axis, which is placed along 
the $x$ coordinate axis, and $b$ being the length of the semi-minor axis. The
parameter $n$ measures the degree of concentration of the bar and $\rho_0$ 
represents the bar central density. 

We also use two further ad-hoc bar potentials, namely a Dehnen's bar type, 
$\Phi_1$, (Dehnen \cite{deh00}):
\begin{equation}
\Phi_1(r,\theta)=-\frac{1}{2}\epsilon v_0^2\cos(2\theta)\left\{{\begin{array}{ll}
\displaystyle 2-\left(\frac{r}{\alpha}\right)^n, & r\le \alpha\rule[-.5cm]{0cm}{1.cm}\\
\displaystyle \left(\frac{\alpha}{r}\right)^n, & r\ge \alpha.\rule[-.5cm]{0cm}{1.cm}
\end{array}}\right., 
\end{equation}
and a Barbanis-Woltjer (BW) bar type, $\Phi_2$, (Barbanis \& Woltjer \cite{bar67}):
\begin{equation}
\Phi_2(r,\theta)=\hat{\epsilon}\sqrt{r}(r_1-r)\cos(2\theta)
\end{equation}
The parameter $\alpha$ is a characteristic length scale of the Dehnen's type 
bar potential, and $v_0$ is a characteristic circular velocity. The parameter 
$\epsilon$ is related to the bar strength. The parameter $r_1$ is a 
characteristic scale length of the BW bar potential and the parameter 
$\hat{\epsilon}$ is related to the bar strength.

\subsection{Equations of motion}
In order to compute the equations of motion, we take into account that the
bar component rotates anti-clockwise with angular velocity ${\bf
\Omega_p}=\Omega_p{\bf z}$, where $\Omega_p$ is a constant pattern
speed\, \footnote{Bold letters denote vector notation. The vector {\bf z} 
is a unit vector.}. The equations of motion in 
this potential in a frame rotating with angular speed ${\bf \Omega_p}$ in 
vector form are
\begin{equation}\label{eq-motvec}
{\bf
\ddot{r}=-\nabla \Phi} -2{\bf (\Omega_p \times \dot{r})-  \Omega_p \times
(\Omega_p\times r)},
\end{equation}
where the terms $-2 {\bf \Omega_p\times \dot{r}}$ and $-{\bf \Omega_p \times
(\Omega_p\times r)}$ represent the Coriolis and the centrifugal
forces, respectively, and ${\bf r}$ is the position vector. Defining an 
effective potential:
\begin{equation}
\Phi_{\hbox{\scriptsize eff}}=\Phi-\frac{1}{2}\Omega_p^2\,(x^2+y^2), 
\end{equation}
Eq. (\ref{eq-motvec}) becomes ${\bf \ddot{r}=-\nabla \Phi_{\hbox{\scriptsize eff}}} -2{\bf (\Omega_p \times \dot{r})},$ and the Jacobi constant is $E_J = \frac{1}{2} {\bf\mid \dot{r}\mid} ^2 + \Phi_{\hbox{\scriptsize eff}},$ which, being constant in time, can be 
considered as the energy in the rotating frame.

\section{Dynamics around $L_1$ and $L_2$}
\label{sec:inv}
\index{Dynamics around $L_1$ and $L_2$}
For our calculations we place ourselves in a frame of reference corotating 
with the bar and place the bar major axis along the $x$ axis. 
In this rotating frame we have five equilibrium points, which, due to the 
similarity with the Restricted Three Body Problem, are called 
Lagrangian points. The points located on the origin of coordinates, namely 
$L_3$, and along the $y$ axis, namely $L_4$ and $L_5$, are linearly stable. 
The ones located symmetrically along the $x$ axis, namely $L_1$ and $L_2$, 
are linearly unstable. Around the equilibrium points there exist families of 
periodic orbits, e.g. around the central equilibrium point the well-known 
$x_1$ family of periodic orbits \cite{con80} that is responsible for the bar 
structure. 

\begin{figure}
\centering
\includegraphics[scale=0.3,angle=-90.]{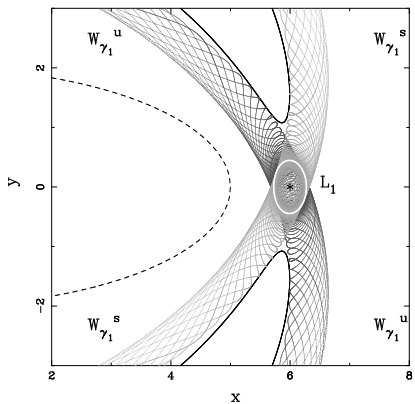}
\caption{Unstable (in dark grey), $W_{\gamma_1}^u$, and stable (in light 
grey), $W_{\gamma_1}^s$, invariant manifolds associated to the periodic 
orbit around $L_1$ (in white solid line). In black solid lines, we plot 
the zero velocity curves for this energy level and the dashed curve shows 
the outline of the bar.}
\label{fig:branches}
\end{figure}

The dynamics around the unstable equilibrium points is described in detail 
in \cite{rom06}; here we give only a brief summary. Around each unstable 
equilibrium point there also exists a family of periodic orbits, known as 
the family of Lyapunov orbits \cite{lya49}. For a given energy level, two 
stable and two unstable sets of asymptotic orbits emanate from the periodic 
orbit, known as the stable and the unstable invariant manifolds, respectively. 
We denote by $W_{\gamma_i}^s$ the stable invariant
manifold associated to the periodic orbit $\gamma$ around the Lagrangian
point $L_i,\, i=1,2$. The stable invariant manifold is the set of orbits 
that tends to the periodic orbit asymptotically. Similarly, we denote by
$W_{\gamma_i}^u$ the unstable invariant manifold associated to the periodic 
orbit $\gamma$ around the Lagrangian point $L_i,\, i=1,2$. The unstable 
invariant manifold is the set of orbits that departs asymptotically from the 
periodic orbit (i.e. orbits that tend to the Lyapunov orbits when the time 
tends to minus infinity) (Fig.~\ref{fig:branches}). Since the 
invariant manifolds extend well beyond the neighbourhood of the equilibrium 
points, they can be responsible for global structures. 

\begin{figure}
\centering
\includegraphics[scale=0.43,angle=-90.]{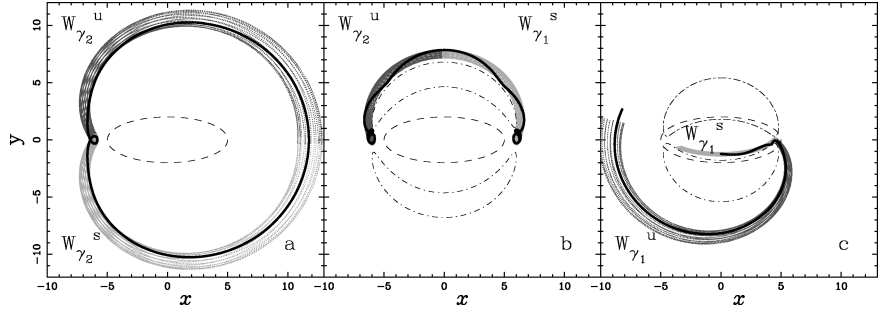}
\caption{Homoclinic {\bf (a)}, heteroclinic {\bf (b)} and transit
{\bf (c)} orbits (black thick lines) in the configuration space. In
dark grey lines, we plot the unstable invariant manifolds associated to the
periodic orbits, while in light grey we plot the corresponding stable invariant
manifolds. In dashed lines, we give the outline of the bar and, in
{\bf (b)} and {\bf (c)}, we plot the zero velocity curves in dot-dashed lines.}
\label{fig:transfer}
\end{figure}

In \cite{rom07}, we give a detailed description of the role invariant 
manifolds play in global structures and, in particular, in the transfer of 
matter. Simply speaking, the transfer of matter is characterised 
by the presence of homoclinic, heteroclinic, and transit orbits. Homoclinic 
orbits correspond to asymptotic trajectories $\psi$ such that $\psi \in 
W_{\gamma_i}^u \cap W_{\gamma_i}^s, \, i=1,2$. That is, they are asymptotic 
orbits that depart from the unstable Lyapunov periodic orbit $\gamma$ around 
$L_i$ and return asymptotically to it (Fig.~\ref{fig:transfer}a). 
Heteroclinic orbits are asymptotic trajectories $\psi^\prime$ such that
$\psi^\prime \in W_{\gamma_i}^u \cap W_{\gamma_j}^s, \, i,j=1,2, \, i\ne j$.
That is, they are asymptotic orbits that depart from the periodic orbit 
$\gamma$ around $L_i$ and asymptotically approach the corresponding 
Lyapunov periodic orbit with the same energy around the Lagrangian point 
at the opposite end of the bar $L_j$, $i\ne j$ (Fig.~\ref{fig:transfer}b). 
We are interested in the homoclinic and heteroclinic orbits corresponding to 
the first intersection of the invariant manifolds with an appropriate surface 
of section. There exist also trajectories that spiral 
out from the region of the unstable  periodic orbit and we refer to them as 
transit orbits (Fig.~\ref{fig:transfer}c). These three types of orbits are 
chaotic orbits since they fill part of the chaotic sea when we plot the 
Poincar\'e surface of section (e.g. the section $(x,\dot x)$ near $L_1$).

\section{Results}
\label{sec:res}
\index{Results}
One of our goals is to check the influence of each main free parameter of 
the models introduced in Sect.~\ref{sec:mod}. In order to do so, 
we make families of models in which only one of the free parameters is 
varied, while the others are kept fixed. Our results in \cite{rom07} show 
that only the bar pattern speed and the bar strength have a considerable 
influence on the shape of the invariant manifolds and, thus, on the 
morphology of the galaxy. Having established this, we perform a 
two-dimensional parameter study for each bar potential and we obtain all 
types of rings and spiral arms.

\begin{figure}
\begin{center}
\includegraphics[angle=-90.,width=0.23\textwidth]{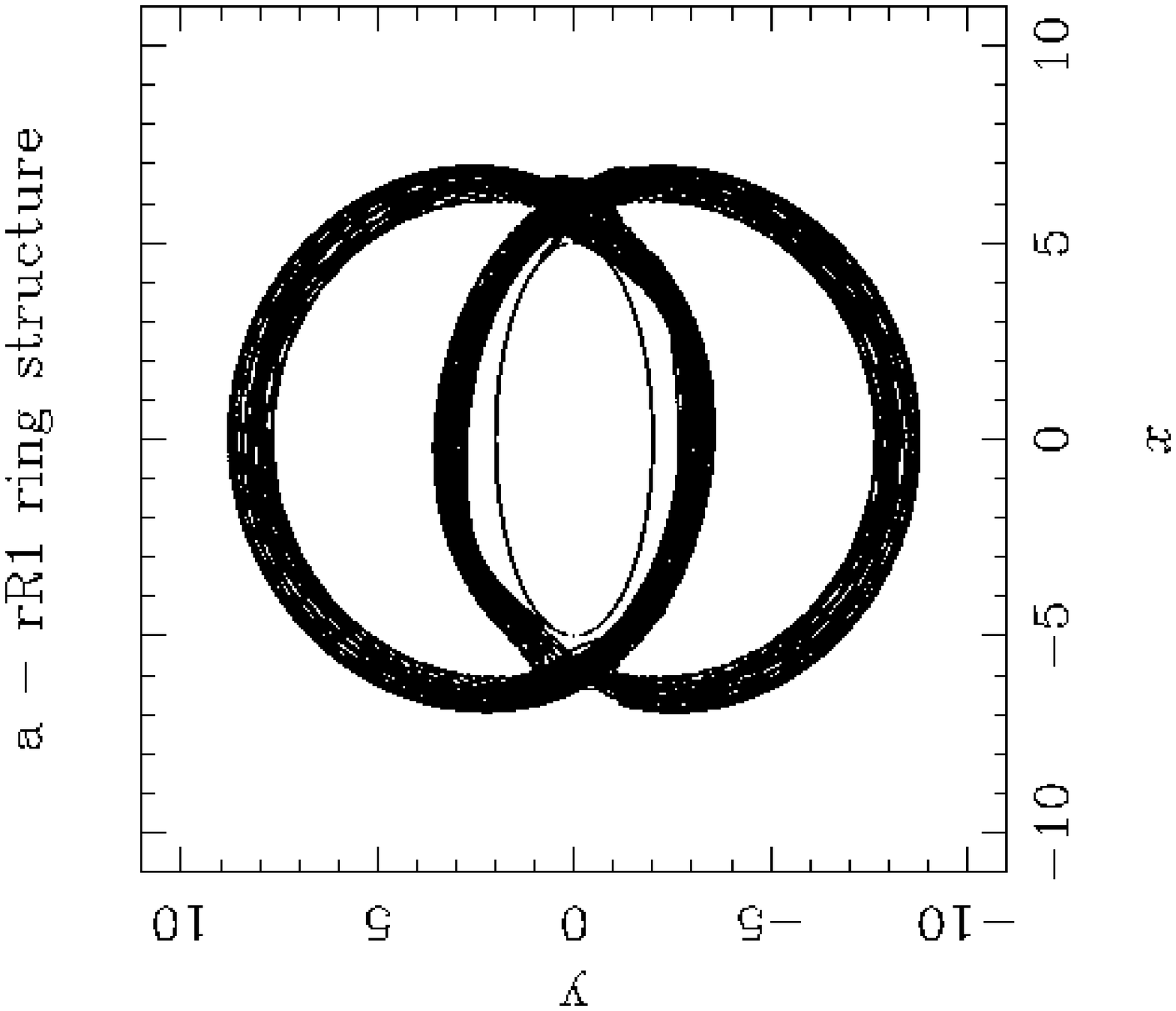}
\includegraphics[angle=-90.,width=0.23\textwidth]{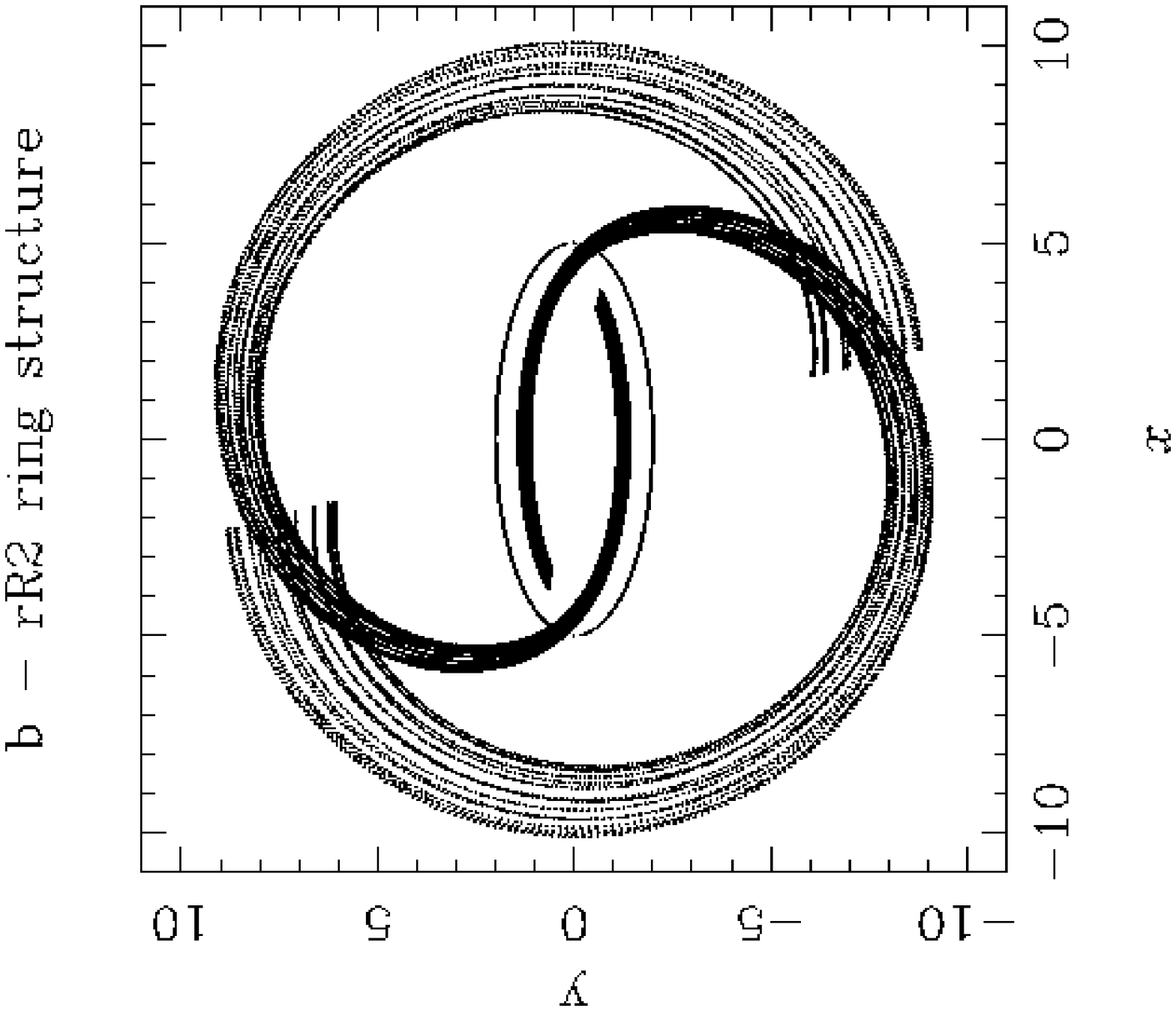}
\includegraphics[angle=-90.,width=0.23\textwidth]{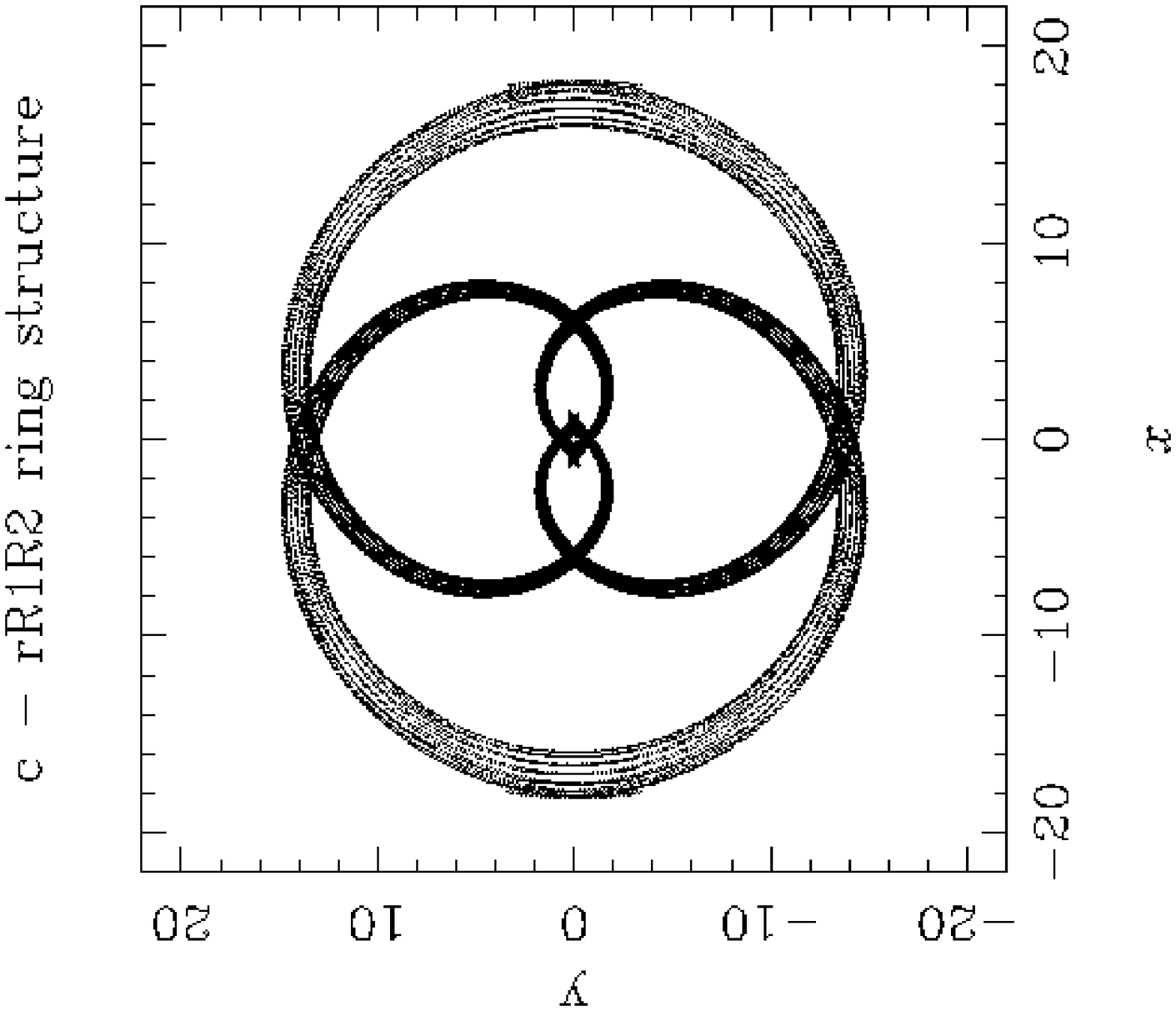}
\includegraphics[angle=-90.,width=0.23\textwidth]{romerogomez_m_fig6.ps}
\end{center}
\caption{Model rings and spiral arm structures. We plot characteristic 
examples of {\bf (a)} $rR_1$ ring structure, {\bf (b)} $rR_2$ ring structure, 
{\bf (c)} $R_1R_2$ ring structure and {\bf (d)} Spiral arms.}
\label{fig:results}
\end{figure}

In Fig.~\ref{fig:results} we show the model rings and the spiral structure
we obtain with our models. We plot the unstable (Fig.~\ref{fig:results}a, b
and d) and the unstable and stable (Fig.~\ref{fig:results}c) invariant
manifolds associated to one of the Lyapunov periodic orbits of the main
family around $L_1$ and $L_2$. Note that we plot the projection of the
invariant manifolds on the configuration space $(x,y)$. Our results show 
that the morphologies obtained depend on dynamical factors, that is, on 
the presence of homoclinic or heteroclinic orbits of the first intersection
of the corresponding invariant manifolds. If heteroclinic orbits exist, 
then the ring of the galaxy is classified as $rR_1$ (Fig.~\ref{fig:results}a). 
The inner branches of the invariant manifolds associated to $\gamma_1$ and 
$\gamma_2$ outline a nearly elliptical inner ring that encircles the bar. 
The outer branches of the same invariant manifolds form an outer ring whose 
principal axis is perpendicular to the bar major axis. If the model has 
neither heteroclinic, nor homoclinic orbits and only transit orbits are 
present, the barred galaxy will present two spiral arms emanating from the 
ends of the bar. The outer branches of the unstable invariant manifolds 
will spiral out from the ends of the bar and they extend azimuthally to
more than $3\pi /2$ (Fig.~\ref{fig:results}d). If the outer branches of the 
unstable invariant manifolds intersect in configuration space with each 
other \footnote{Note that they cannot intersect in phase space.}, then they 
form the characteristic shape of $R_2$ rings (Fig.~\ref{fig:results}b). 
That is, the trajectories outline an outer ring whose major axis is parallel 
to the bar major axis. The last possibility is if only homoclinic orbits 
exist. In this case, the inner branches of the invariant manifolds form an 
inner ring, while the outer branches outline both types of outer rings, thus 
the barred galaxy presents an $R_1R_2$ ring morphology 
(Fig.~\ref{fig:results}c).

\begin{figure}
\centering
\includegraphics[scale=0.45]{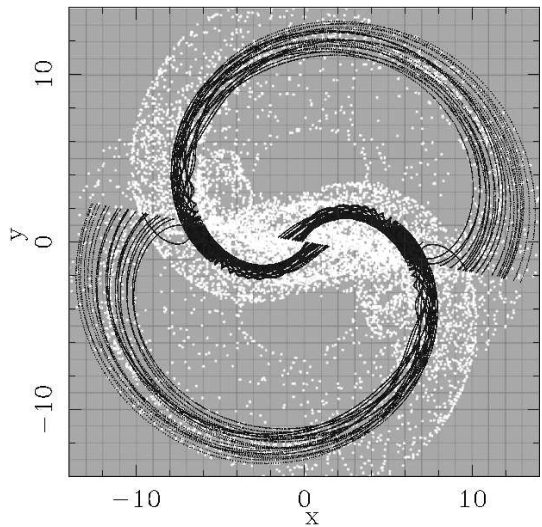}
\caption{Over-plot of the results obtained with the response simulation (white 
dots) and the invariant manifolds (black lines) in a model with spiral arms.}
\label{fig:response}
\end{figure}

We also study the response of an axisymmetric component to a 
bar perturbation. We use the same axisymmetric potential and the same bar 
potential as in our models and the bar is introduced gradually, to avoid 
transients. Once the bar has reached its maximum amplitude, we consider a 
snapshot of the response simulation and we compare its morphology to the 
corresponding structure we obtain with our models. In 
Fig.~\ref{fig:response} we show the results for the spiral arms case, by
over-plotting the selected snapshot with our model. The white points 
represent the particle positions of the response study and the black lines 
are the unstable invariant manifolds. Note that the two match perfectly.

We compare our results with observational data (E. Athanassoula, M. 
Romero-G\'omez, J.J. Masdemont, C. Garc\'{i}a-G\'omez, in preparation) and 
we find good agreement. Regarding the photometry, the density profiles across 
radial cuts in rings and spiral arms agree with the ones obtained from 
observations. The velocities along the ring also show that these are only a 
small perturbation of the circular velocity. 

\section*{Acknowledgements}
MRG acknowledges her fellowship ``Becario MAE-AECI''.

%
%

%
%

\index{References}


\printindex
\end{document}